\documentclass[journal=mamobx,manuscript=communication]{achemso}
\usepackage{graphicx,color}

\title{Easy orientation of diblock copolymers on self-assembled monolayers using UV irradiation}
\author{Pang-Hung Liu}
\affiliation{CEA,IRAMIS,LIONS, CEA-Saclay, F-91191 Gif-sur-Yvette Cedex, France}

\author{Patrick Guenoun}
\affiliation{CEA,IRAMIS,LIONS, CEA-Saclay, F-91191 Gif-sur-Yvette Cedex, France}

\author{Jean Daillant}
\affiliation{CEA,IRAMIS,LIONS, CEA-Saclay, F-91191 Gif-sur-Yvette Cedex, France}
\email{Jean.Daillant@cea.fr}

\date{\today}

\begin{document}
\maketitle

\abstract{
A simple method based on UV/ozone treatment is proposed to control the surface energy of dense 
grafted silane layers for orientating block copolymer mesophases. 
Our method allows one to tune the surface energy down to a fraction of a mN/m. 
We show that related to the surface, perpendicular orientation of a lamellar phase of a PS-PMMA diblock 
copolymer (neutral surface) is obtained for a critical surface energy of 
23.9-25.7 mN/m. Perpendicular cylinders are obtained for 24.6mN/m and
parallel cylinders for 26.8 mN/m.
}
\vskip 1cm
The control of the surface energy of self-assembled monolayers (SAMs) for orientating diblock 
copolymer mesophases has been achieved using different methods.
The goal is usually to obtain a neutral surface, that is a surface exhibiting similar 
interfacial energies with both blocks of the copolymer in order to promote perpendicular orientation.
Historically, the first method to be used was to spin coat random copolymers 
made of the same monomers on the substrate \cite{kellogg96}, 
as different surface energies can be achieved by varying the ratios of monomers in synthesis.
Further, a major improvement was to modify the random copolymer at one end with 
a chemical group which could be grafted on the surface in order to prevent 
the random copolymer to dewet or diffuse in the block copolymer \cite{mansky97,huang98}.
Besides of this surface specific chemistry, cross-linking on the random copolymer 
has also recently been used to increase the film stability \cite{ryu07}. 
More recently, the thickness dependence of the orientation has been investigated 
\cite{ham08} and it has been shown that the formation of perpendicular domains 
on a random copolymer brush needs to be viewed in terms of the equilibration of the block copolymer and 
the random copolymer in the presence of each other and not simply in terms of interfacial energy \cite{han08}.\\
A different strategy has been to use self-assembled monolayers (SAM) like silanes. 
However, though it has been shown that 3-({\it p}-methoxyphenyl)propyltrichlorosilane
could provide neutral surfaces for PS-PMMA (polystyrene-poly(methyl methacrylate)) on a silicon wafer \cite{park07},
such monolayers usually don't have the right surface energy and 
different methods have been used in order to control this energy. 
It has for example been shown that incomplete silane monolayers could also be used \cite{niemz06}. 
Whereas this method is extremely versatile, it leads to a necessarily heterogeneous 
film which might prevent control at very small scales.
Another possibility is irradiation of SAMs with synchrotron soft X-rays \cite{kim00,peters00}, or treatment 
with CO$_{2}$ plasma \cite{Delorme2006612}.
Oxidation through ultraviolet (UV) radiation has also been
recently used to produce wettability gradients \cite{smith03}. 
In this paper we show that UV irradiation indeed provides a versatile tool 
to precisely tune the surface energy of a SAM. We show in particular that the orientation 
of PS-PMMA lamellar and cylinder phases can be controlled using this method.\\

Si wafers (p-type, boron dopped, 250 $\mu m$ thick) 
were first cleaned by sonication in purified Millipore 
water (resistivity 18 M$\Omega$.cm), 1:1 water/ethanol mixture, 
chloroform and heptane, followed by piranha treatment 
(1/3 v/v of 30\% H$_{2}$O$_{2}$/98\% H$_{2}$SO$_{4}$) at $80\,^{\circ}{\rm C}$ and exposure to UV under oxygen atmosphere for 30 min. 
The cleaned wafers were rinsed with purified water, dried with nitrogen and silanized in a 2 mM solution of octadecyltrichlorosilane (OTS) in heptane for 1 day. 
The wafers were then sonicated in chloroform and purified water and dried with nitrogen before UV/ozone treatment.\\
The silanized wafers were then exposed to UV light (185 and 254 nm) in an oxygen filled chamber 
at a distance of 4 cm from the lamp, 
for specific time periods, rinsed with purified water and chloroform and dried with nitrogen.\\

The OTS layers were first characterized using x-ray reflectivity using a 
Siemens powder diffractometer D5000 operated 
with a home-made software. The Cu K$_\alpha$ 
is first collimated using 50-$\mu$m slits. A graphite monochromator
is placed after the sample in front of the NaI 
scintillator detector. The home-made software allows one to record rocking curves at each point of
the reflectivity curve in order to subtract the background.
Best fit of the reflectivity curve to a model describing the OTS layer as a single 
slab of constant electron density gives a substrate roughness of 0.4 nm, an OTS layer 
thickness of 2.45 nm and an OTS layer -air roughness of 0.65 nm in good agreement
with Ref. \cite{Tidswell90}. 
After 6 min UV ozone treatment, the thickness of the OTS layer decreases to 2.27 nm 
whereas its roughness increases to 0.785 nm, indicating changes in the top surface structure
and composition. Attempts to add a specific surface layer to the model did not lead
to significantly better fits.\\

\begin{figure}
\includegraphics{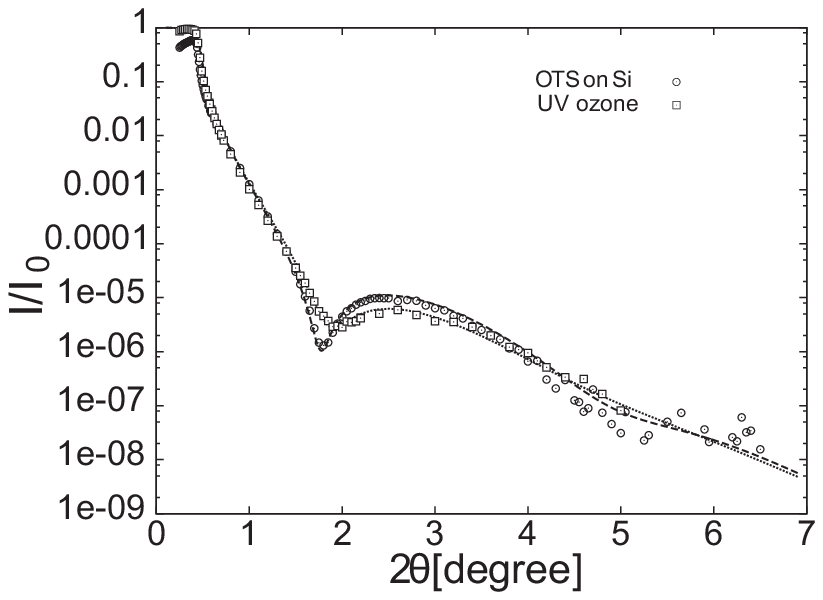}
\caption{ Reflectivity results of a typical silane layer (circle) and a 6-min UV ozone treated silane layer with the calculated curves (dash line and dot line respectively)}
\label{figure1}
\end{figure}

\begin{figure}
\includegraphics{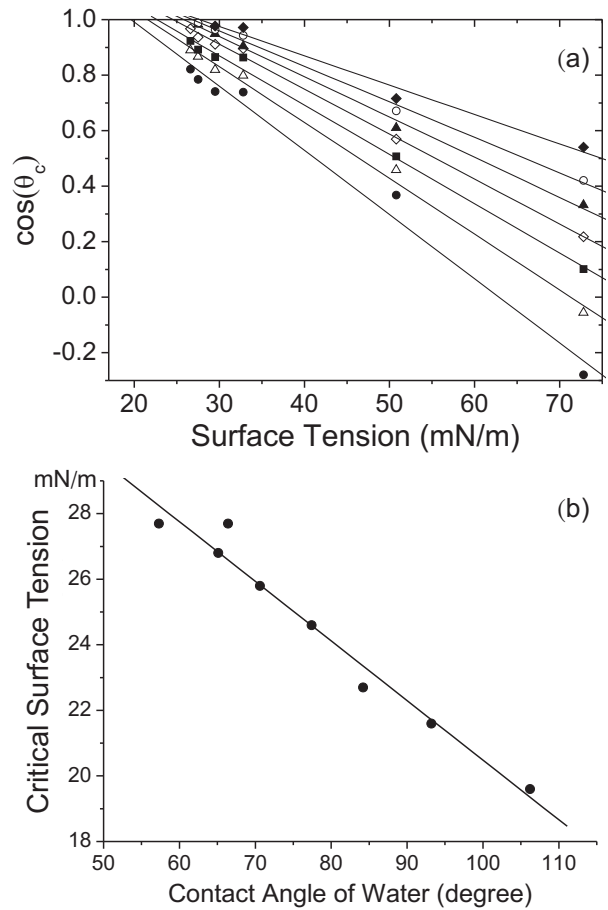}
\caption{(a) Zisman plots, cosine of the contact angle $\cos \theta_c$ as a 
function of the surface tension for 
tetradecane, hexadecane, squalane, bicyclohexyl, diiodomethane and purified water
for silanated wafers treated with different UV/ozone time periods:
native wafer (filled circles), 3 min
(open triangles), 4 min
(filled squares), 5 min
(open diamonds), 6 min
(filled triangles), 7 min
(open circles), 9 min
(filled diamonds). (b) Critical surface tensions as determined by the intercepts of the curves  as a function of water contact angle chosen as a sensitive characteristic parameter of the surface.
}

\label{figure2}
\end{figure}

The surface energy of the native and oxidized silane layers were also carefully characterized.
The surface energy of solid surfaces is usually characterized through so-called Zisman plots 
which consist in plotting the cosine of the contact angle $\theta$ of an homologous series of liquids as a function of the liquid surface 
tension $\gamma$ of the liquid (\ref{figure2}(a))\cite{degennes85,zisman64}. 
The intercept at $\cos \theta$=1 
gives the critical surface tension of the surface $\gamma_C$ which indicates whether a given liquid
will wet ($\gamma < \gamma_C$) or not ($\gamma > \gamma_C$) the substrate.
In principle, non-polar homologous liquids like alkanes should be used, but in practice this also works
very well even with water.\cite{widom97}
From fitting the Zisman plots for tetradecane, hexadecane, squalane, bicyclohexyl, diiodomethane and purified water  (\ref{figure2}(a)),
we obtain the critical surface tensions shown in \ref{figure2}(b).The critical surface energies
range from 19.6 mN/m for the native surface layer in good agreement with literature \cite{brzoska94}, 
to 27.7 mN/m after 8 minutes UV irradiation, demonstrating that our method allows a very precise control 
of the surface energy, down to a fraction of a mN/m. In order to characterize the surface independently of our irradiation time (a parameter which may vary from an apparatus to another), we report in \ref{figure1}(b) the critical surface tension as a function of the water contact angle. This  parameter is very sensitive to the oxidation state of the surface. A very smooth and linear variation is then obtained over a range of angles very similar to the one found by Han et al.\cite{han08}..\\

Diblock copolymers of  PS$_{52K}$-{\it b}-PMMA$_{52K}$ (PDI: 1.09) 
of symetric composition and  PS$_{46K}$-{\it b}-PMMA$_{21K}$ (PDI: 1.09) of  asymetric composition were purchased 
from {\it Polymer Source Inc}. 
PS$_{52K}$-{\it b}-PMMA$_{52K}$ exhibits a lamellar phase of period $L_0 = 49 \rm{nm}$ 
in the bulk whereas PS$_{46K}$-{\it b}-PMMA$_{21K}$ 
exhibits a  phase of PMMA cylinders (of about 36 nm for the center to center spacing) in PS.
1 wt\% solutions of PS$_{52K}$-{\it b}-PMMA$_{52K}$ and PS$_{46K}$-{\it b}-PMMA$_{21K}$ 
in toluene were spin-coated onto silanized silicon wafers treated with UV/ozone at 2000 and 2500 rpm to produce copolymer films 
with thicknesses {\it ca.} 34 nm and 31 nm, respectively. 
Subsequently, the samples were annealed in a vacuum oven of pressure less than 3 kPa
at $170\,^{\circ}{\rm C}$ for 1 day. \\
 
Atomic force microscopy (AFM, Digital Instruments, Nanoscope V) was employed in tapping mode for 
imaging PS-PMMA films at room temperature. Phase images allow one to easily distinguish PS (dark) from PMMA (bright) domains \cite{magonov97}. 
After spin coating, due to low surface energies, 
no copolymer was observed on non-treated ($\gamma_C=19.6 \rm{ mN/m}$) and 3-min UV/ozone treated ($\gamma_C=21.6 \rm{mN/m}$)samples.
For the samples with water contact angles of $87\,^{\circ}$ and $84\,^{\circ}$, dewetting can be observed under optical microscopy and AFM (inserts of \ref{figure3}(a,f)).
As shown in \ref{figure3}(a-e)
perpendicular orientation of the lamellar phase of the 
PS$_{52K}$-{\it b}-PMMA$_{52K}$ copolymer is obtained on the samples with $\gamma_C$ from 23.9 to 25.7 mN/m.
Its period is {\it ca.}50 nm as expected from bulk studies. 
\ref{figure3}(f-j) show the different morphogies obtained after UV/ozone treatment
for the PS$_{46K}$-{\it b}-PMMA$_{21K}$ copolymer films. 
First, on the sample with $\gamma_C=22.7 \rm{mN/m}$, dewetting of copolymer is observed after annealing (\ref{figure3}(f)). 
Perfect perpendicular orientation of the cylinders is obtained after 5 min UV/ozone treatment
(\ref{figure3}(g), $\gamma_C$=24.6 \rm{mN/m}). From the FFT of AFM images, the average distance between them is
33 nm.
After 6 min (\ref{figure3}(h), $\gamma_C$=25.8mN/m), the phase contrast is obviously
much reduced and the orientation is lost.
After 7 min treatment, elongated stuctures are observed (\ref{figure3}(i), $\gamma_C$=26.8mN/m).
The structure gives a spacing of 39 nm, which is close to the value of $33 \rm{nm} \times 2/\sqrt{3} $.
This indicates that they should be the same cylindrical structure observed on 5-min treated sample 
but lie flat on the surface.
Finally, after 8 and 9 min treatment (\ref{figure3}(j), $\gamma_C$=27.7 mN/m) this orientation is lost again.\\

In this paper, we have demonstrated that UV/ozone oxidation of OTS layers provides a simple
and versatile way of controlling the orientation of block copolymers through a precise control
of the surface energy. As already noticed in the literature, the window for the perpendicular orientation
of the lamellar phase is wider than for the cylindrical phase \cite{ham08}.  
According to literature, the surface tensions of PS and PMMA at 170$^\circ C$ are 
$\gamma_{\rm PS} \approx 29.7-29.9 mN/m$ $\gamma_{\rm PMMA} \approx 29.9-31 mN/m$ respectively
\cite{wu70,mansky97,chee98}.
If we assume that the neutral surface would have a surface energy of $\approx 29.9 mN/m$, then this 
would be $\approx 5mN/m$ more than the corresponding critical surface tension of $24.5 mN/m$.
According to \cite{wu70}, there is a good agreement between surface energy and critical surface 
tension for PMMA but a large discrepancy for PS and polyethylene, for which the critical 
surface tension is also found to be $\approx 4mN/m$ smaller than the $20^\circ C$ surface energy. Whereas the OTS 
chains probably expose more methyl groups to the surface (which would also explain the lower
critical surface tension with respect to polyethylene surface tension), this shows that such a difference is not unrealistic.

{\bf Acknowledgements:} Pang-Hung Liu gratefully acknowledges support form the ``Chimtronique'' 
program of CEA. The authors also gratefully thank O. Tach\'e et C. Blot for support
during the x-ray experiments.

\bibliography{panghung}

\newpage

\begin{table} [p]
\centering
\renewcommand{\arraystretch}{1.2}
\tabcolsep=15pt

\begin{tabular}{llll}
\hline
Sample & $\theta_{C,water} (^{\circ})^a$ & $\gamma_C (mN/m)^b$ & description$^c$ \\
\hline  
C1  & 84   & 22.7  &  \\
C2  & 77   & 24.6  & $C_\bot$ \\
C3  & 71   & 25.8  &  \\
C4  & 65   & 26.8  & $C_{//}$ \\
C5  & 57   & 27.7  &  \\
L1  & 87   & 22.7$^d$  & dewetting \\
L2  & 80   & 23.9$^d$  & $L_{//}$ \\
L3  & 77   & 24.5$^d$  & $L_{//}$ \\
L4  & 70   & 25.7$^d$  & $L_{//}$ \\
L5  & 62   & 27.1$^d$  & dewetting \\
\hline
\multicolumn{4}{l}{$^a$Contact angles of water in degrees.}  \\
\multicolumn{4}{l}{$^b$Critical surface energy values in mN/m.} \\
\multicolumn{4}{l}{$^c$The phases or morphology  observed in AFM images.} \\
\multicolumn{4}{l}{$^d$Values calculated from linear interpolation of $\theta_{C,water}$.} \\
\end{tabular}
\caption{Results of contact angles, critical surface energies and the phases observed in AFM images.}
\label{table1}
\end{table}

\end{document}